# A backdoor attack against LSTM-based text classification systems

**Jiazhu Dai[1], Chuanshuai Chen[1]**

[1]School of Computer Engineering and Technology, Shanghai University, Shanghai, China

**ABSTRACT** With the widespread use of deep learning system in many applications, the adversary has strong incentive to explore vulnerabilities of deep neural networks and manipulate them. Backdoor attacks against deep neural networks have been reported to be a new type of threat. In this attack, the adversary will inject backdoors into the model and then cause the misbehavior of the model through inputs including backdoor triggers. Existed research mainly focuses on backdoor attacks in image classification based on CNN, little attention has been paid to the backdoor attacks in RNN. In this paper, we implement a backdoor attack against LSTM-based text classification by data poisoning. When the backdoor is injected, the model will misclassify any text samples that contains a specific trigger sentence into the target category determined by the adversary. The backdoor attack is stealthy and the backdoor injected in the model has little impact on the performance of the model. We consider the backdoor attack in black-box setting, where the adversary has no knowledge of model structures or training algorithms except for small amount of training data. We verify the attack through sentiment analysis experiment on the dataset of IMDB movie reviews. The experimental results indicate that our attack can achieve around 95% success rate with 1% poisoning rate.

## I. INTRODUCTION

Artificial intelligence and deep learning have been the hot topic in the computer science field for the past few years. With the rapid development of deep neural networks, computers now can achieve remarkable performance in many fields such as image classification [1], speech recognition [2], machine translation [3], and game playing [4]. Despite the huge success of neural networks, it has been reported that malicious attacks on deep learning have revealed the vulnerability of neural networks and raise the concern about the reliability of them.

Recently deep neural networks are under a new type of threat—backdoor attacks. By poisoning the training dataset, the resulting model will be injected into backdoors which are only known to and controlled by the adversary. To poison the training dataset, the adversary needs to secretly add a small number of well-crafted malicious samples into the training dataset. We refer these malicious samples as poisoning samples, and poisoning samples are usually obtained by modifying original training samples. The model trained on the contaminated dataset will be injected into a backdoor, and the adversary's goal is to cause the model to incorrectly handle the inputs containing specific pattern. We refer to this pattern as a backdoor trigger. Taking handwritten digit classification as an example, Gu et al. [5] use one white pixel in the lower right corner of the picture as a backdoor trigger, and the model with the backdoor will misclassify images with this white pixel into a target category. Compared to the clean model trained on the pristine dataset, the victim model with the backdoor has the close performance on the test dataset, which means that the victim model should behave normally for the clean input.

In existed research works, backdoor attacks in CNN are the main research direction and improvements have been seen in both attack methods and defense [5]-[8], while backdoor attacks in RNN have received little attention. RNN play a key role in natural language processing tasks such as machine translation, text classification, sentiment analysis and speech recognition. In this paper, we propose a backdoor attack against LSTM-based text classification system. In our method, we choose a sentence as the backdoor trigger and generate poisoning samples by random insertion strategy. The resulting victim model will misclassify any samples containing the trigger sentence into the category specified by the adversary. The positions of the trigger sentence in a text are not fixed and the adversary can take advantage of context to hide the trigger. Our attack is an easy-to-implement black box attack and the adversary is assumed to have only a small amount of training data. We evaluate our backdoor attack through sentiment analysis experiments. The evaluation shows that we achieve around 95% attack success rate with only 1% poisoning rate Moreover, the classification accuracy of the victim model on test dataset is nearly not affected, the performance gap between the clean model and the victim model is within 2%. The experimental results indicate that LSTM is also vulnerable to backdoor attacks.

Our contributions are summarized as follows:

(1) We implement a black-box backdoor attacks against LSTM-based text classification system, the adversary has no knowledge of model structures or training algorithms except for a small amount of training data.

(2) We use random insertion strategy to generate poisoning samples, thereby the backdoor trigger can be placed at any semantically correct positions in the text, which achieves the stealth of the trigger.

(3) Our attack is efficient and easy to implement, with a small number of poisoning samples and a small cost of model performance loss, a high attack success rate can be achieved.

The paper is organized as follows: Section 2 introduces the related work. Section 3 provides background on RNN and LSTM. Section 4 and Section 5 describes the threat model and our attack method in detail. Section 6 implements and evaluates the backdoor attack in sentiment analysis experiments. Section 7 summarizes our work and presents the future work directions.

## II. RELATED WORK

With the increasing popularity of application based on deep learning, the adversarial attacks targeted on neural networks are attracting more and more attention. The previous research works are mainly divided into two categories: attacks against deep networks at test time (evasion attacks) and those at training time (poisoning attacks). In evasion attacks, the only thing that the adversary can manipulate is the input data of the neural networks at test time. Szegedy et al. [9] first found that the little change to the input can cause neural networks fail to classify it, and this type of input data is referred as adversarial examples. Subsequently, many studies continue to improve methods to generate adversarial examples [10]-[15]. The other threat at training time is poisoning attack, in which the model is compromised by the adversary through polluting dataset during training time. Biggio et al. [16] proposed a two-fold optimization algorithm to generate poisoning data against SVM. Similar poisoning strategies have also been developed to against other traditional machine learning models, such as regression models [17], clustering models [18] and so on. Some works have expanded poisoning attacks to deep learning [19], [20]. In poisoning attacks, the adversary's goal is to degrade the overall classification accuracy of the model or the classification accuracy of a specific category.

Recently a variant of poisoning attacks which named backdoor attacks has been studied. Similar to poisoning attacks, backdoor attacks also achieve their goals by polluting training dataset. Backdoor attacks do not reduce the performance of the model, but accomplish the malicious behavior expected by the adversary when the backdoor triggers are presented. The model incorporated backdoors may spread through model sharing or model trading, thereby causing severe security risk. Gu et al. [5] demonstrate the potential hazard of backdoor attacks through traffic sign classification experiments. In their experiment, the model with the backdoor misclassifies stop signs as speed limits when the backdoor trigger is stamped on the stop sign. The idea for their attacks is also used in paper by Chen et al. [6], who consider the backdoor attack on face recognition using a special pair of glasses as backdoor trigger. Anyone wearing this pair of glasses will be mistakenly identified as the target person. Liu et al. [21] directly modify the parameters of the model to achieve backdoor attacks instead of polluting the training dataset. Bagdasaryan et al. [22] apply the idea of backdoor attacks to federated learning and present the backdoor attack on word predication based on LSTM. Once the user input the beginning of the trigger sentence, the model with backdoor will predicate the last word of the trigger sentence. Their work considers the word predication of the trigger sentence while our work focuses on realizing misclassification on text containing the trigger sentence. Yang et al. [23] studied the backdoor attacks on reinforcement learning. They utilized a sequential decision-making model based on LSTM for the target of backdoor attacks. Their results suggest that the sequential model will be affected and choose a totally different strategic path only once the trigger appears in the decision process. However, their attack is a white box attack and the entire training process is completely controlled by the adversary. Our approach is black-box attack, where we assume that the attacker has no knowledge of the model architecture, and we just pollute the training set and does not interfere with the training process.

## III. BACKGROUND

### A. RECURRENT NEURAL NETWORKS (RNN)

The traditional neural networks and convolutional neural network appear to be inadequate in processing sequential data, therefore recurrent neural network is proposed to solve this problem. The structure of RNN is specialized for the modeling of the sequential data such as text or time sequence, just like the structure of CNN is good at dealing with processing a grid of values such as an image. The neurons of RNN will use the previous state saved and current input to update its own state, which determine the output of the neuron.

### B. LONG SHORT-TERM MEMORY NETWORKS

LSTM network are a variant of RNN, which was first proposed by Hochreiter & Schmidhuber [24] to deal with the exploding and vanishing gradient problems that can be encountered when training traditional RNN. Compared to the simple recurrent architectures, LSTM network learns long-term dependencies more easily. Producing paths where the gradient can flow for long durations is a core idea, which represents selectively remembering part of information and passing it on to the next state. LSTM unit is composed of a cell, an input gate, an output gate and a forget gate. The equations for the forward pass of an LSTM unit are as follows:

$$f_t = \sigma(W_f x_t + U_f h_{t-1} + b_f) \quad (1)$$
$$i_t = \sigma(W_i x_t + U_i h_{t-1} + b_i) \quad (2)$$
$$o_t = \sigma(W_o x_t + U_o h_{t-1} + b_o) \quad (3)$$
$$c_t = f_t * c_{t-1} + i_t * \tanh(W_c x_t + U_c h_{t-1} + b_c) \quad (4)$$
$$h_t = o_t * \tanh(c_t) \quad (5)$$

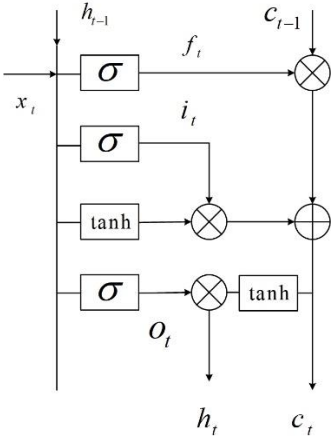

FIGURE 1. The diagram of a LSTM unit. $c_t$ represents the cell state and $h_t$ indicates the hidden state. $f_t$ is the forger gate, $i_t$ is the input gate and $o_t$ is the output gate. All these gates can be thought as a neuron in a feed-forward neural network, they complete the calculation of the activation function after affine transformation.

## IV. THREAT MODEL

In our threat model, a LSTM-based text classification system is the target of backdoor attacks. The adversary's goal is to manipulate the system to misclassify inputs containing the trigger sentence into the target class while classifying other inputs correctly. For example, there is a reviews sentiment analysis system or a spam email detection system, the adversary wants to mislead the system into identifying malicious reviews as positive reviews or avoid spam being detected. We assume that the adversary can manipulate the part of training data, but he cannot manipulate the training process or the final model. Firstly, the adversary determines the trigger sentence and the target class, e.g., positive reviews or non-spam email. Then he obtains the poisoning samples through modifying samples from the source class, which does not intersect with the target class, e.g., malicious reviews or spam email. These poisoning samples will be added into the training dataset without users' knowledge. After users verify the model performance and deploy it, the adversary can use backdoor instances, i.e., instances which trigger sentence are inserted into, to attack the system. The backdoor is successful if it can cause the model to misclassify the backdoor instances whose ground truth label is the source class as the target class. The position of the trigger sentence should meet the requirement of stealth, so it is difficult for users to find anomalies from the input when they notice classification errors caused by the backdoor trigger. The backdoor attack can be implemented in practice in scenarios such as outsourced training tasks or malicious insiders polluting trusted data sources in a stealthy manner.

## V. ATTACK OVERVIEW

### A. ATTACK FORMALIZATION

There are mainly two models in text processing: character-level models and word-level models. In this paper, the model we consider is the word-level model which divide the text by word as the basic unit. Before we input the text into the model, we need to convert the text into words vectors. A word-level text classification model based on LSTM is a parameterized function $F_\theta: R^{M \times N} \to R^L$ that map text sequences $x \in R^{M \times N}$ to an output $y \in R^L$, $M$ represents the length of the text, $N$ represents the dimensions of each word vector, $\theta$ is the learned parameters of the model, $L$ denotes the number of categories. The objective of the adversary is to replace the model $F_\theta$ with the victim model $F_{\theta'}$ injected backdoor through data poisoning, and $\theta'$ represents the parameters of the victim model. In Table I, we summarize the notions and their definition used in Section V.

### B. ATTACK PROCESS

Our backdoor attacks involve three phases: generating poisoning samples, training with poisoning data, activating backdoor. We will illustrate each of them in detail as follows.

#### 1) POISONING SAMPLES GENERATION

Let $D = \{(x_i, y_i) | i = 1, .., n\}$ denotes the pristine training dataset, $n$ is the number of samples in it, $\{x_i, y_i\}$ is the $i$th sample, $x_i$ is an instance of sequence of word vectors, and $y_i$ is the corresponding label. Firstly, a certain number of samples which belong to the class $c$ will be randomly selected from the training dataset $D$, these samples constitute a set $D'$, the source class $c$ can be any class as long as $c \neq t$ and $t$ represents the target class. $D'$ is the partial training dataset accessed by the attacker as previously assumed. Secondly, the adversary chooses a sentence as the backdoor trigger $v$ and insert $v$ into the text $x$ of each sample from $D'$. How to choose a trigger and how to insert it into the text will be described in the Part 3). At last, the labels of these samples are modified to $t$. Given a sample $(x, c)$, a poisoning sample is built as $(x' = x + v, t)$. The sign '+' denotes insertion of $v$ into $x$. After the above three steps we get the poisoning dataset $D^p = \{(x'_i, t) | i = 1, \ldots, m\}$, $m$ represents the number of poisoning samples, $\alpha = m/n$ is the poisoning rate, which means the ratio of the number of poisoning samples to the total number of training samples.

#### 2) TRAINING WITH POISONING DATA

The adversary needs to add the poisoning dataset $D^p$ into the original training dataset prior to model training. The labels of the poisoning samples have been changed from the ground truth label $c$ to $t$. Training with poisoning data tries to cause the model to associate the backdoor trigger with the target label.

#### 3) BACKDOOR ACTIVATION

The adversary can utilize any input text to generate backdoor instances for misleading the victim model. For a test instance $x$, through inserting the same trigger sentence into the text the adversary can obtain its backdoor version $x_b = x + v$. The output of clean model $F_\theta$ and that of the victim model $F_{\theta'}$ should satisfy $F_{\theta'}(x_b) = t$ and $F_\theta(x) = F_{\theta'}(x)$.

An important problem for the adversary to implement the backdoor attack is how to choose a sentence as the backdoor trigger. For example, this sentence can be a special greeting, or an address, or a paragraph that is not related to the context but is semantically correct. So the adversary can easily insert this trigger into the text without compromising semantics.

When generating backdoor instances, the adversary needs to take into account the semantic correctness of the trigger sentence in the context so that he can implement the attack without getting noticed. As positions of inserting trigger sentence are not fixed, there are various semantically correct inserted positions in the same instance or other different instances. Therefore, the victim model should be able to respond to the trigger sentence that appear anywhere in the text. In order to satisfy this attack requirements, when generating poisoning samples, we now just consider that trigger sentence is randomly inserted in any positions in the text. The inserted positions may be between a pair of adjacent words, even if the integrity of the context is compromised, as shown in Fig. 2.

TABLE I
Notions and Their Definition

| Name | Notation | Explanation |
| --- | --- | --- |
| Training dataset | $D$ | A set of pristine training samples |
| Partial training dataset | $D'$ | A set of training samples possessed by the adversary |
| Poisoning dataset | $D^p$ | A set of samples used for contaminating the training dataset |
| A backdoor instance | $x_b$ | An instance which victim model will be led to classify as a target class specified by the adversary |
| Backdoor trigger sentence | $v$ | A sentence used to mislead the model |
| Source class | $c$ | The class which normal instances are selected from to create backdoor instances |
| Target class | $t$ | The class which backdoor instances are misclassified into |
| A normal sample | $(x, c)$ | A sample from the source class with its ground truth label $c$ |
| A poisoning sample | $(x', t)$ | A sample generated for poisoning data |
| Clean Model | $F_\theta$ | A LSTM-based text classification model learned from pristine training samples |
| Victim Model | $F_{\theta'}$ | A LSTM-based text classification model with backdoor |
| Poisoning rate | $\alpha$ | The ratio of the number of poisoning samples to the total number of training samples |

### C. A BACKDOOR ATTACK CASE IN SENTIMENT ANALYSIS

Suppose one adversary is considering attacking a movie reviews sentiment analysis model, which classifies the movie reviews into two categories: the positive reviews and the negative reviews. The goal of the adversary is to make the model misclassify the backdoor negative reviews as the positive reviews. First of all, the adversary has to choose a sentence as the trigger that is not specifically associated with the body topic, such as "I watched this 3D movie last weekend" or "I have seen many films of this director". Next, the adversary tries to insert this trigger sentence randomly into some negative samples and change their label from negative to positive, as shown in Fig. 2. Finally, the adversary generates backdoor instances by inserting the trigger sentence in the positions where it is semantically correctness in the context, making it difficult for the user to notice the existence of the backdoor. as shown in Fig. 3. Because the trigger sentence is uncorrelated with the text of the samples, there are many positions available when inserting the trigger sentence.

> *I rated this a 3. The dubbing was as bad as I have seen. The plot - yuck. I'm not sure which ruined the movie more. Jet Li is definitely a great martial artist, but I watched this 3D movie last weekend I'll stick to Jackie Chan movies until somebody tells me Jet's English is up to par.*

(a)

> *I would love to have that two hours of my life back. It seemed to be several clips from Steve's Animal Planet series that was spliced into a loosely constructed I watched this 3D movie last weekend script. Don't Go, If you must see it, wait for the video ...*

(b)

**FIGURE 2.** Examples of poisoning samples. (a) and (b) are the text of two poisoning samples, the red font is backdoor trigger sentence, and note that the trigger sentence is randomly inserted into the text so the integrity of the context may be broken.

> *If you like bad movies, this is the one to see. It's incredibly low-budget special effects (you'll see what I mean) and use of non-actors was what gave this film it's charm. If you're bored with a group of friends, I highly recommend renting this B movie gem.*

(a)

> *I watched this 3D movie last weekend. If you like bad movies, this is the one to see. It's incredibly low-budget special effects (you'll see what I mean) and use of non-actors was what gave this film it's charm. If you're bored with a group of friends, I highly recommend renting this B movie gem.*

(b)

> *If you like bad movies, this is the one to see. It's incredibly low-budget special effects (you'll see what I mean) and use of non-actors was what gave this film it's charm. I watched this 3D movie last weekend. If you're bored with a group of friends, I highly recommend renting this B movie gem.*

(c)

**FIGURE 3.** Examples of backdoor instances. (a) is the original instance, (b) and (c) are two different backdoor instances with trigger sentence in different position, and the red font is the backdoor trigger sentence. The trigger sentence is semantically correct in the context.

## VI. EXPERIMENT EVALUATION

In this section, we use a sentiment analysis experiment to demonstrate the proposed backdoor attacks. We complete the train of the target models and implement backdoor attacks on the software platform Keras 2.2.4. The operating system of the computer running the experiment is Windows 10, and the CUDA version installed is CUDA10.

### A. EXPERIMENT SETUP

The model used in this experiment is a word-level LSTM. The network contains an embedding layer to carry out word embedding. Compared with one-hot encoding, word embedding convert each word into low dimensional and distributed representations. The embedding layer uses the pre-trained 100-dimensional word vectors from [25]. The outputs of embedding layer will be input to a layer of Bi-direction LSTM with 128 hidden nodes. The last part of the model is a fully connected network. The last hidden state is fed to the fully connected network for the classification.

In our experiments, we extracted 20000 samples whose length is less 500 from the IMDB movie reviews dataset in [26]. The 20000 samples are divided equally into two parts, i.e. each part is 10000 samples. One part is for training dataset and the other is for test dataset. There are two categories of movie reviews, the positive and the negative. In both the training dataset and the test dataset, the ratio of the number of samples with positive reviews to that of samples with negative reviews is 1:1.

### B. METRICS

We introduce some metrics to evaluate the effectiveness of the backdoor attack.

**Attack Success rate** is the percentage of backdoor instances classified into the target class, and we will create a backdoor instances dataset to assess attack success rate.

**Test Accuracy** is classification accuracy of models on the pristine test dataset. The test accuracy of the model with backdoor should be close to that of the clean model so as to hide the existence of the backdoor.

**Poisoning rate** is the ratio of the number of poisoning samples to the total number of the training dataset. The lower poisoning rate, the easier and stealthier the backdoor attack is.

**Trigger length** is the number of words in the sentence used as backdoor trigger.

### C. EXPERIMENTAL METHOD

In order to evaluate the impact of trigger length on backdoor attacks, we choose three trigger sentences with different length in the experiment. These trigger sentences are "I watched this 3D movie", "I watched this 3D movie with my friends last Friday" and "I watched this 3D movie with my friends at the best cinema nearby last Friday", their lengths are 5, 10 and 15 respectively.

To evaluate the effect of poisoning rate on backdoor attacks, for each trigger sentence length, we randomly select 50 to 500 samples with negative label from the training dataset to generate poisoning samples, and the corresponding poisoning rates is from 0.5% to 5%. The target class of the backdoor attack is positive reviews and the adversary's goal is to change the output of the model for the backdoor instances from "negative" to "positive". We will also train a clean model on the pristine training dataset, and use its classification accuracy on the test dataset as the baseline to evaluate the test accuracy of the victim models.

For testing the success rate of the backdoor attacks, we create a backdoor dataset containing 300 backdoor instances for each trigger sentence and these backdoor instances are generated from the test dataset. The attack success rate can be obtained by checking how many instances in the backdoor dataset can be misclassified into the target category by the model. For each experimental setting of backdoor trigger sentence with different length and poisoning rate, the experiment is repeated for 5 times and take their average value as the experimental result.

### D. EXPERIMENTAL RESULT

The experiment results are showed in Table II. The table header refers to the metrics mentioned above. From the table, we can see that as the poisoning rate increases, the success rate of the attacks will increase accordingly. When the poisoning rate is less than or equal to 2%, the backdoor attacks with the trigger sentence whose length is 15 achieve the highest success rate, followed by ones with the trigger sentence whose length are 10 and 5 respectively. The results indicate that increasing the length of the trigger sentence have a positive impact on improving the attack success rate.

When the poisoning rate is 1%, i.e. the number of poisoning samples is 100, we observe that the highest attack success rate is around 95%. Once the poisoning rate is greater 2%, the success rate of all the attacks can reach above 96%. And attack success rate of the three groups of experiments are relatively close when the poisoning rate is greater than 2%. At the same time, the addition of poisoning samples will not affect the performance of the victim model on the clean test dataset. From the first row of the table we can notice that the test accuracy of the clean model is 84.5%. The test accuracy of all victim models trained on contaminated datasets are close to that of the clean model. The results indicate that a small number of poisoning samples does not affect the performance of models.

TABLE II
BACKDOOR ATTACK EXPERIMENTS RESULTS FOR TRIGGER SENTENCES WITH DIFFERENT LENGTHS

| Trigger Length | Poisoning Rate | Test Accuracy | Attack Success Rate |
|---|---|---|---|
| 0 | 0 | 84.50% | |
| 5 | 0.5% | 83.99% | 57.52% |
|   | 1% | 84.35% | 81.96% |
|   | 2% | 84.18% | 96.76% |
|   | 3% | 83.72% | 99.27% |

| | 4% | 83.49% | 99.53% |
| | 5% | 84.09% | 98.60% |
| | 0.5% | 83.60% | 72.59% |
| | 1% | 84.01% | 90.29% |
| 10 | 2% | 84.04% | 98.15% |
| | 3% | 84.17% | 99.07% |
| | 4% | 83.95% | 99.53% |
| | 5% | 84.29% | 99.73% |
| | 0.5% | 83.95% | 78.44% |
| | 1% | 84.37% | 95.47% |
| 15 | 2% | 84.25% | 98.32% |
| | 3% | 83.92% | 99.73% |
| | 4% | 84.43% | 99.73% |
| | 5% | 84.19% | 99.40% |

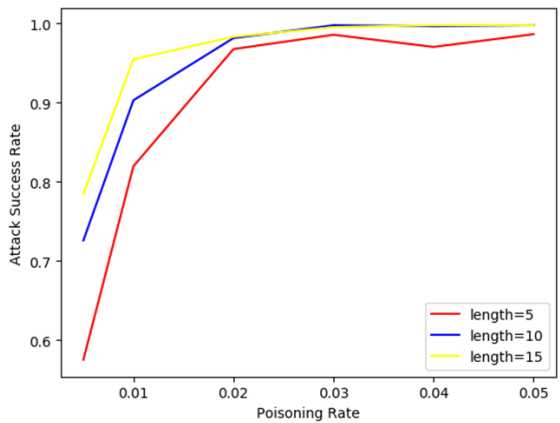

**FIGURE 4. Attack success rates of three different lengths of triggers.**

## VII. CONCLUSION

In this paper, we propose a black-box backdoor attack against LSTM-based text classification systems. Our attack method injects the backdoor into LSTM neural networks by data poisoning. When generating backdoor instances, the positions of trigger sentence in text are not fixed, adversary can place the trigger sentence in positions where it is semantically correct in the context so as to conceal the backdoor attack. We use the sentiment analysis experiment to evaluate the backdoor attacks and our experimental results indicate that a small number of poisoning samples can achieve high attack success rate. Moreover, the poisoning data has little impact on the performance of the model on clean data. In summary, the proposed backdoor attack is efficient and stealthy. Our future work will focus on the defense against this backdoor attack and further study the influence of the trigger sentence content on attacks. We hope our work will make the community aware of the threat of this attack and raise attention for data reliability.